\documentclass{ifacconf}

\usepackage{graphicx}      
\usepackage{natbib,url}        

\newcommand{\R}{{\mathcal R}}

\newcommand{\tr}{{\rm tr}}
\newcommand{\sk}{{\rm sk}}
\newcommand{\sym}{{\rm sym}}

\newcommand{\N}{{\mathcal{N}}}

\newcommand{\V}{{\mathcal V}}
\newcommand{\E}{{\mathcal E}}

\begin{document}
\begin{frontmatter}

\title{Vision-Based Cooperative Estimation of Averaged 3D Target Pose 
under Imperfect Visibility} 


\author[First]{Takeshi Hatanaka} 
\author[First]{Takayuki Nishi} 
\author[First]{Masayuki Fujita}

\address[First]{Department of Mechanical and Control Engineering Tokyo Institute of Technology, 
   2-12-1 S5-26 Ookayama Meguro-ku, Tokyo 152-8550, Japan (e-mail: hatanaka@ctrl.titech.ac.jp).}

\begin{abstract}
This paper investigates vision-based cooperative estimation
of a 3D target object pose for visual sensor networks.
In our previous works, we presented an 
estimation mechanism called 
networked visual motion observer achieving averaging of
local pose estimates in real time.
This paper extends the mechanism
so that it works even in the presence of cameras not viewing the target
due to the limited view angles and obstructions
in order to fully take advantage of the networked vision system.
Then, we analyze the averaging performance attained by the proposed mechanism
and clarify a relation between the feedback gains in the 
algorithm and the performance.
Finally, we demonstrate the effectiveness of the algorithm
through simulation.
\end{abstract}


\end{frontmatter}


\section{Introduction}

Driven by technological innovations of smart wearable vision
cameras, a networked vision system 
consisting of spatially distributed smart cameras
emerges as a new challenging application field of the visual 
feedback control and estimation (\cite{SKSD1,TrVi:11}).
The vision system called {\it visual sensor network}
brings in some potential advantages over a
single camera system such as:
(i) accurate estimation by integrating rich information, 
(ii) tolerance against
obstructions, misdetection in image processing
and sensor failures
and (iii) wide vision and elimination of blind areas
by fusing images of a scene from a variety of viewpoints. 
Due to their nature, the visual sensor networks are 
expected as a component of sustainable infrastructures.

Fusion of control techniques and visual information
has a long history, which is well summarized by
\cite{CH_06,CH_07,vision}.
Among a variety of estimation/control problems 
addressed in the literature, this paper investigates 
a vision-based estimation problem of
3D target object motion as in \cite{AH_SCL,aswin1,TCST07}.
While most of the above works consider estimation
by a single or centralized vision system,
we consider a {\it cooperative estimation} problem
for visual sensor networks.
In particular, we confine our focus to a 3D pose estimation problem 
of a moving target object addressed by \cite{TCST07},
where the authors present a real-time vision-based observer
called {\it visual motion observer}.
Namely, we investigate cooperative estimation
of a target object pose via distributed processing.

Cooperative estimation for sensor networks has been addressed
e.g. in \cite{Olf:07,FYL_CDC06}. 
The main objective of these researches is averaging the local 
measurements or local estimates among sensors 
in a distributed fashion to improve estimation accuracy.
For this purpose, most of the
works utilize the consensus protocol (\cite{OFM_IEEE07})
in the update procedure of the local estimates.
However, the consensus protocol is not applicable to the full
3D pose estimation problem as pointed out by \cite{TrVi:11},
since the object's pose takes values in a non-Euclidean space.

Meanwhile, \cite{TrVi:11,SS_SIAM09} present a 
distributed averaging algorithm on matrix manifolds.
However, applying them to cooperative estimation
requires a lot of averaging iterations at each update of
the estimate and hence they cannot deal with the case
where the target motion is not slow.
To overcome the problem, the authors presented a cooperative estimation
mechanism called {\it networked visual motion observer}
achieving distributed estimation of an object pose in real time
(\cite{CDC11,TAC13}) by using a pose synchronization techniques in
\cite{TAC12}.
However, \cite{CDC11,TAC13} assume that all the cameras capture
the target object, which may spoil the advantage (iii)
of the first paragraph of this section.
Though running the algorithm only among the cameras 
viewing the target and broadcasting the estimates to the other cameras
is an option, it is desirable
to share an estimate without changing procedures of each camera
in order to avoid such complicated task switches depending on the situation.

In this paper, we thus present a novel estimation mechanism which works
in the presence of cameras not capturing the target
due to limited view angles and obstructions.
Then, we analyze the averaging performance attained by the proposed mechanism
and clarify a relation between the tuning gains and the averaging performance.
There, we prove that the conclusion of \cite{CDC11,TAC13}
under the assumption of perfect visibility is also valid
even in the case of imperfect visibility.
Moreover, we demonstrate the effectiveness of the presented algorithm
through simulation.


\section{Problem Statement}
\label{sec:problem}

\begin{figure}
\begin{center}
\begin{minipage}{4.2cm}
\includegraphics[width=4.2cm]{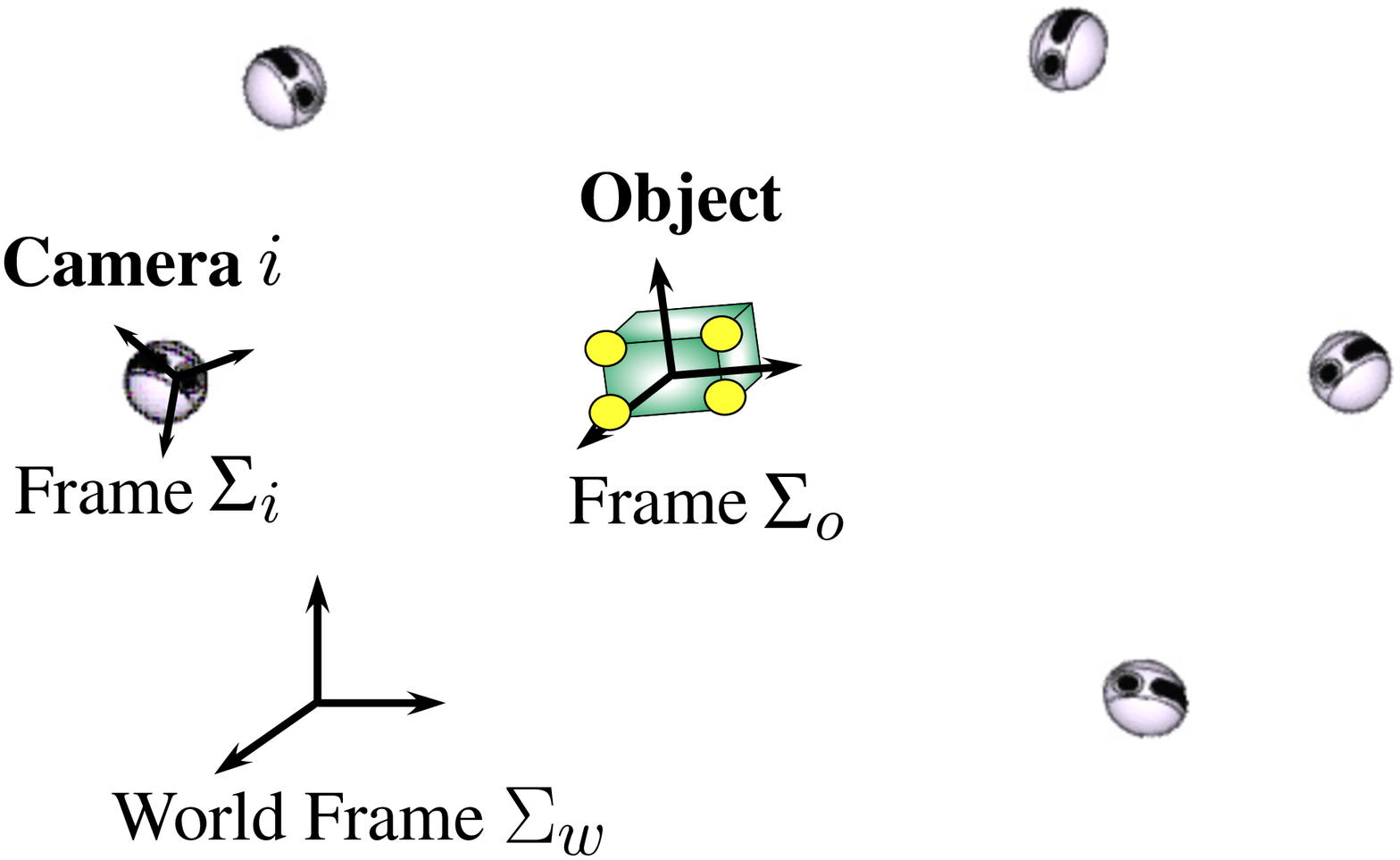}
\end{minipage}
\hspace{.2cm}
\begin{minipage}{4.2cm}
\includegraphics[width=4.2cm]{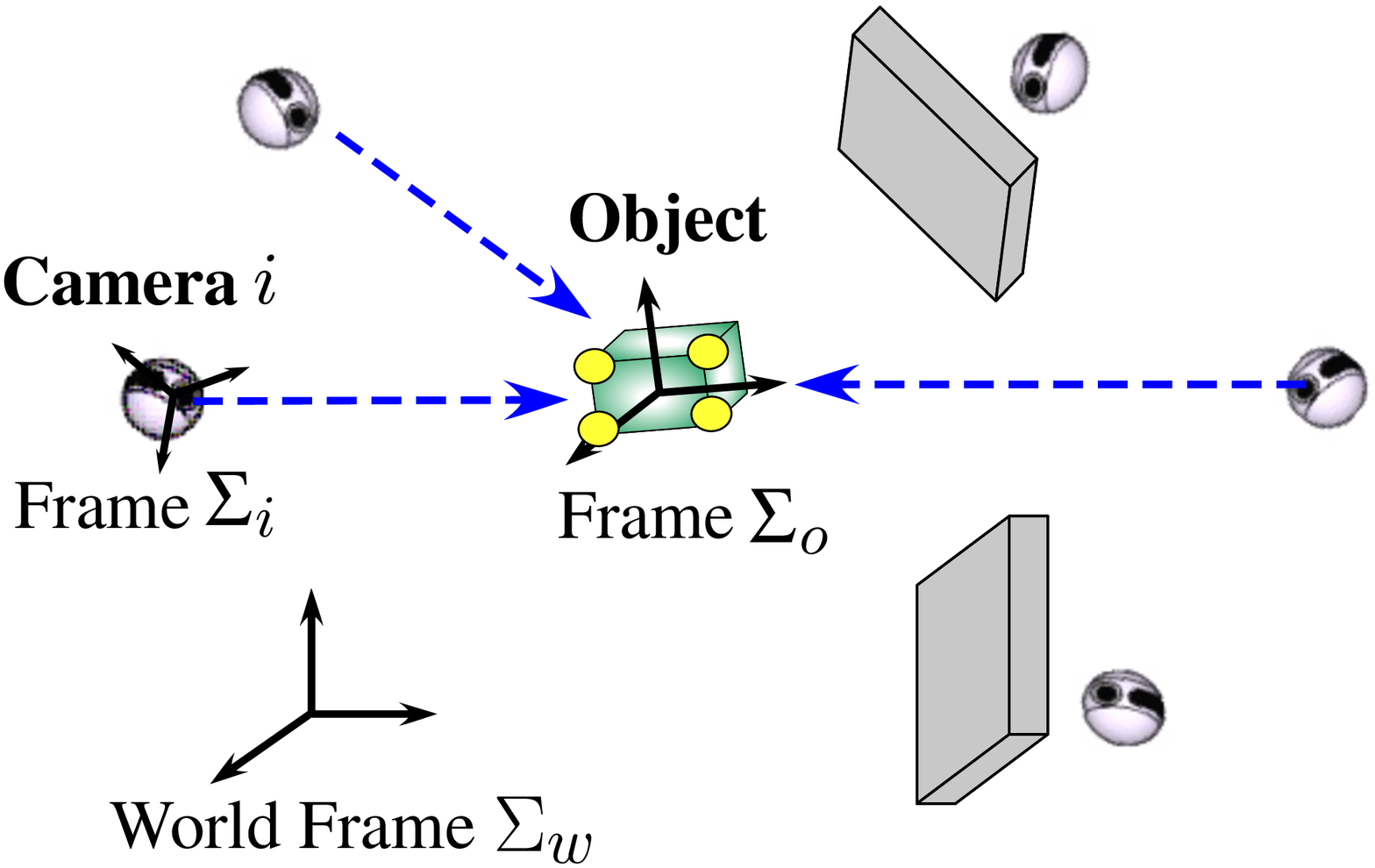}
\end{minipage}
\caption{Situation under consideration}
\label{fig:visible}
\end{center}
\end{figure}

\subsection{Situation under Consideration}

In this paper, we consider the situation where
there are $n$ cameras $\V := \{1, \cdots, n\}$ with
communication and computation capability and a single target object 
in 3 dimensional space as in the left figure
of Fig. \ref{fig:visible}.
Let the world frame, the $i$-th camera frame and the object frame 
be denoted by $\Sigma_w$, $\Sigma_i$ and $\Sigma_{o}$, respectively.
The objective of the networked vision system is to estimate
the 3D pose of the object from visual measurements.
Although the targets are possibly multiple in a practical situation,
we confine our focus only to estimation of a single target
since multiple objects case can be handled
by just applying parallely the procedure 
for a single object to each object.

Unlike \cite{CDC11,TAC13},
all the vision cameras are assumed to have visible region
and some cameras do not capture the target object
as depicted in the right figure of Fig. \ref{fig:visible}.
Let us now denote the subset of all vision cameras 
viewing the target at time $t$ by $\V_f(t) \subseteq \V$ and the rest of the cameras by $\V_f^-(t) \subseteq \V$.

\begin{figure}
\begin{center}
\includegraphics[width=8cm]{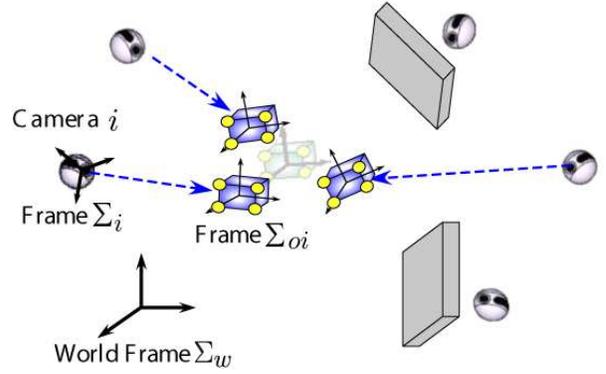}
\caption{Sensing under uncertainties}
\label{fig:situation}
\end{center}
\end{figure}

Suppose that the pose consistent with the visual measurement of 
each camera $i \in \V_f(t)$
differs from camera to camera due to incomplete localization
and parametric uncertainties of the cameras 
as depicted in Fig.~\ref{fig:situation}.
Then, the fictitious target with the 
pose consistent with the $i$-th camera's visual measurement
is denoted by $o_i,\ i \in {\mathcal V}_f(t)$ and
its frame is by $\Sigma_{o_i},\ i \in \V_f(t)$.
Under such a situation, averaging the contaminated poses is
a way to improve estimation accuracy (\cite{Olf:07,TrVi:11}).
In this paper, we thus address estimation of an average pose
of objects $\{o_i\}_{i \in \V}$ in a distributed fashion.

\subsection{Relative Rigid Body Motion}

The position vector and the rotation matrix from $i$-th camera frame $\Sigma_i$ to the world frame $\Sigma_w$ are denoted by $p_{wi} \in \R^3$ and $e^{\hat{\xi}_{wi}\theta_{wi}} \in SO(3) := \{ R \in \R^{3 \times 3} 
| R^TR = RR^T = I_3,\ \det(R) = +1 \}$.
The vector $\xi_{wi} \in \R^3$ specifies the rotation axis and $\theta_{wi} \in \R$ is the rotation angle.
We use $\xi\theta_{wi}$ to denote $\xi_{wi}\theta_{wi}$.
The notation $'\wedge'$ is the operator such that $\hat{a}b = a \times b$, $a, b \in \R^3$ for the vector cross-product $\times$,
i.e. $\hat{a}$ is a $3 \times 3$ skew-symmetric matrix.
The notation $'\vee'$ denotes the inverse operator to $'\wedge'$.

The pair of the position $p_{wi}$ and the orientation 
$e^{\hat{\xi}\theta_{wi}}$ denoted by $g_{wi} = (p_{wi}, 
e^{\hat{\xi}\theta_{wi}}) \in SE(3) := \R^3 \times SO(3)$ is called the 
pose of camera $i$ relative to the world frame $\Sigma_w$.
Similarly, we denote by $g_{wo_i} = (p_{wo_i}, e^{\hat{\xi}\theta_{wo_i}}) \in SE(3)$ the pose of object $o_i$ relative to the world frame $\Sigma_w$.
We also define the body velocity of camera $i$ relative to the world frame $\Sigma_w$ as $V^b_{wi} = (v_{wi}, \omega_{wi}) \in \R^6$, where $v_{wi}$ and $\omega_{wi}$ respectively represent the linear and angular velocities of the origin of $\Sigma_i$ relative to $\Sigma_w$.
Similarly, object $o_i$'s body velocity relative to $\Sigma_w$ is denoted by $V^b_{wo_i} = (v_{wo_i}, \omega_{wo_i}) \in \R^6$.

Throughout this paper, we use the following homogeneous representation of $g = (p, e^{\hat{\xi}\theta}) \in SE(3)$ and $V^b = (v, \omega)$.
\begin{eqnarray} 
g = \left[
\begin{array}{cc}
	e^{\hat{\xi}\theta} & p \\
	0 & 1 \\
\end{array}
\right], \hat{V}^b = \left[
\begin{array}{cc}
	\hat{\omega} & v \\
	0 & 0 \\
\end{array}
\right].
\nonumber
\end{eqnarray}
Then, the body velocities $V^b_{wi}$ and $V^b_{wo_i}$ are simply given by $\hat{V}^b_{wi} = g_{wi}^{-1}\dot{g}_{wi}$ and $\hat{V}^b_{wo_i} = g_{wo_i}^{-1}\dot{g}_{wo_i}$.

Let $g_{io_i} = (p_{io_i}, e^{\hat{\xi}\theta_{io_i}}) \in SE(3)$ be the pose of $\Sigma_{o_i}$ relative to $\Sigma_i$.
Then, it is known that $g_{io_i}$ can be represented as $g_{io_i} = g_{wi}^{-1}g_{wo_i}$.
By using the body velocities $V^b_{wi}$ and $V^b_{wo_i}$, the motion of the relative pose $g_{io_i}$ is written as
\begin{eqnarray} \label{eq:RRBM}
\dot{g}_{io_i} = - \hat{V}^b_{wi}g_{io_i} + g_{io_i} \hat{V}^b_{wo_i}
\end{eqnarray}
(\cite{vision}).
(\ref{eq:RRBM}) is called relative rigid body motion.

\subsection{Visual Measurement}

In this subsection, we define visual measurements of each vision camera 
$i\in \V_f(t)$ which is available for estimation.
Unlike \cite{CDC11,TAC13}, all the cameras in $\V_f^-(t)$
obtain no measurement. 
Now, we assume (i) all cameras are pinhole-type cameras, (ii) each target 
object has $m\ (m\geq 4)$ feature points and (iii) each camera can extract them from the vision data.
The position vectors of object $o_i$'s $l$-th feature point relative to $\Sigma_{o_i}$ and $\Sigma_i$ are denoted by $p_{o_il} \in \R^3$ and $p_{il} \in \R^3$ respectively.
Using a transformation of the coordinates, we have $p_{il} = g_{io_i}p_{o_il}$, where $p_{o_il}$ and $p_{il}$ should be regarded with a slight abuse of notation as $[ p_{o_il}^T\  1 ]^T$ and $[ p_{il}^T\  1 ]^T$.

\begin{figure}
\begin{center}
\includegraphics[width=8cm]{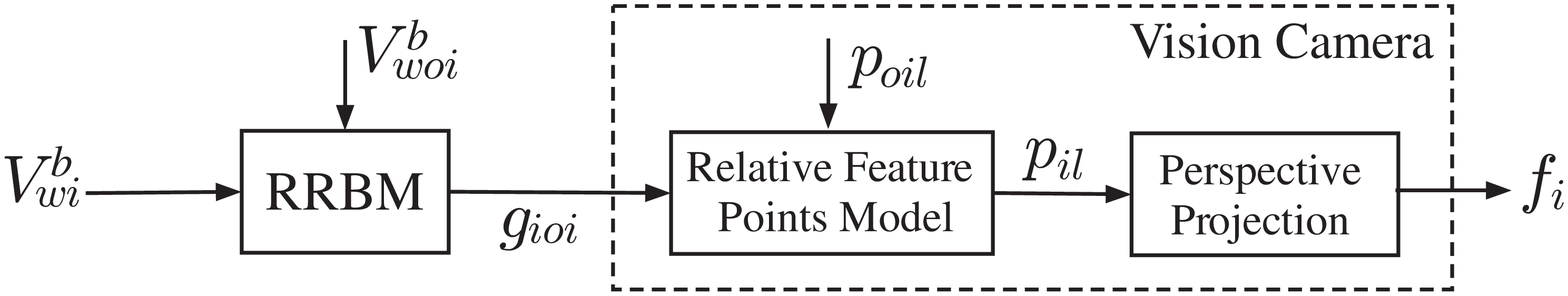}
\caption{Relative rigid body motion with camera model}
\label{fig:plant}
\end{center}
\end{figure}

Let the $m$ feature points of object $o_i$ on the image plane coordinate be the measurement $f_i$ of camera $i$, which is given by the perspective projection (\cite{vision})  as
\begin{eqnarray} \label{eq:visual_mea}
f_i &:=& \left[
\begin{array}{ccc}
	f_{i1}^T & \cdots & f_{im}^T
\end{array}
\right]^T \in \R^{2m}, 
f_{il} = \frac{\lambda_i}{z_{il}}\left[
\begin{array}{c}
	x_{il} \\
y_{il}
\end{array}
\right].
\end{eqnarray}
with a focal length $\lambda_i$, where
$p_{il} = \left[
\begin{array}{ccc}
	x_{il} & y_{il} & z_{il}
\end{array}
\right]^T$.
In this paper, we assume that each camera $i\in \V$ knows the location of feature points $p_{o_il} \in \R^3$.
Then, the visual measurement $f_i$ depends only on the relative pose $g_{io_i}$ from (\ref{eq:visual_mea}) and $p_{il} = g_{io_i}p_{o_il}$.
Fig. \ref{fig:plant} shows the block diagram of the relative rigid body motion (\ref{eq:RRBM})
with the camera model (\ref{eq:visual_mea}), where RRBM
is the acronym of Relative Rigid Body Motion.

\subsection{Communication Model}
The cameras have communication capability with the neighboring cameras and form a network.
The communication is modeled by a graph $G = (\V, \E)$, $\E \subset \V \times \V$.
Namely, camera $i$ can get information from $j$ if $(j,i) \in \E$.
We also define the neighbor set $\N_i$ of camera $i \in \V$ as
\begin{eqnarray} \label{eq:ni}
\N_i := \{j \in \V | (j,i) \in \E \}.
\end{eqnarray}
In this paper, we employ the following assumption on the graph $G$.
\begin{assum} \label{assum:graph}
The communication graph $G$ is fixed, undirected and connected.
\end{assum}

We also introduce some additional notations.
Let ${\mathcal T}(i_0)$ be the set of all 
spanning trees over $G$ with a root $i_0 \in \V$
and we consider an element $G_T = (\V, \E_T)\in {\mathcal T}(i_0)$.
Let the path from $i_0$ to node $i\in \V$ along with $G_T$
be denoted by $P_{G_T}(i) = (v_0,\cdots, v_{d_{G_T}(i)}),\ v_0 = i_0,\ v_{d_{G_T}(i)} = 
i,\ (v_l,v_{l+1}) \in \E_T\ {\forall l}\in \{0, \cdots,d_{G_T}(i)-1\}$,
where $d_{G_T}(i)$ is the length of the path $P_{G_T}(i)$.
We also define 
\[
\Delta_{G_T}(E;i) =\left\{
\begin{array}{ll}
1,&\mbox{if the path } P_{G_T}(i) \mbox{ includes edge } E\\
0,&\mbox{otherwise}
\end{array}
\right.
\]
for any $E \in \E_T$ and
\begin{eqnarray}
 W &:=& \min_{i_0 \in \V} D(i_0),\ D(i_0) := \min_{G_T \in {\mathcal T}(i_0)}\tilde{D}(G_T),\label{eqn:defD}\\
\tilde{D}(G_T) &:=& \max_{E\in \E_T}\sum_{i\in \V}\Delta_{G_T}(E;i)d_{G_T}(i).
\nonumber
\end{eqnarray}
The meaning of these notations are given in \cite{TAC13}.

\subsection{Average on SE(3)}

The objective of this paper is to present a cooperative estimation 
mechanism for the visual sensor networks producing an estimate close to 
an average of $\{g_{io_j}\}_{i \in \V,\ j \in \V_f}$, $g_{io_j} := g_{wi}^{-1}g_{wo_j}$ even in
the presence of vision cameras not capturing the target. 

Let us now introduce the following mean $g^*$ on $SE(3)$ (\cite{Moa:02})
as an average of target poses $\{g_{wo_j} \}_{j \in \V_f}$.
\begin{eqnarray} \label{eq:g_ave}
g^* = (p^*, e^{\hat{\xi}\theta^*}) := \arg \min_{g \in SE(3)} \sum_{j \in \V_f}\psi(g^{-1}\tilde{g}_{wo_j}) 
\end{eqnarray}
where $\psi$ is defined for any $g = (p, e^{\hat{\xi}\theta}) \in SE(3)$ as
\begin{eqnarray} \label{eq:psi}
\psi(g) &:=& \frac{1}{2}\| I_4 - g \|^2_F = \frac{1}{2}\|p \|^2 + \phi(e^{\hat{\xi}\theta}), \\
\label{eq:phi}
\phi(e^{\hat{\xi}\theta}) &:=& \frac{1}{2}\| I_3 - e^{\hat{\xi}\theta} \|^2_F = \tr(I_3 - e^{\hat{\xi}\theta}) 
\end{eqnarray}
and $\|M \|_F$ is the Frobenius norm of matrix $M$.
Hereafter, we also use the notation $g^*_i = (p^*_i, e^{\hat{\xi}\theta^*_i}) := g_{wi}^{-1}g^*$.

\section{Networked Visual Motion Observer}

In this section, we introduce a cooperative estimation mechanism 
originally presented by \cite{CDC11}. 
Here, we assume that the 
relative poses $g_{ij} = g_{wi}^{-1}g_{wj}$ w.r.t neighbors $j 
\in \N_i$ are available for each camera $i\in \V$.

\subsection{Review of Previous Works}

We first prepare a model of the rigid body motion (\ref{eq:RRBM}) as 
\begin{eqnarray} \label{eq:EsRRBM}
\dot{\bar{g}}_i = - \hat{V}_{wi}^b \bar{g}_i + \bar{g}_i \hat{u}_{ei},
\end{eqnarray}
where $\bar{g}_i = (\bar{p}_i, e^{\hat{\bar{\xi}}\bar{\theta}_i})$ is 
the estimate of the average $g^*_i = g_{wi}^{-1}g^*$.
The input $u_{ei} = (v_{uei}, \omega_{uei})$ is to be designed so that 
$\bar{g}_i$ approaches $g^*_i$.
Once $\bar{g}_i$ is determined,
the estimated visual measurement $\bar{f}_i$ is computed by (\ref{eq:visual_mea}).

Let us now define the error $g_{ei} := \bar{g}_i^{-1}g_{io_i}$ between the estimate $\bar{g}_i$ and the relative pose $g_{io_i}$ and its vector representation $e_{ei} := E_R(g_{ei})$ with
\begin{eqnarray} \label{eq:ER}
E_R(g) &:=& \left[
\begin{array}{cc}
	p^T & e_R^T(e^{\hat{\xi}\theta}) \\
\end{array}
\right]^T, \ 
e_R(e^{\hat{\xi}\theta}) := \sk(e^{\hat{\xi}\theta})^{\vee}, \nonumber \\
\sk(e^{\hat{\xi}\theta}) &:=& \frac{1}{2}(e^{\hat{\xi}\theta} - e^{-\hat{\xi}\theta}).
\end{eqnarray}
It is shown by \cite{TCST07} that if the number of feature 
points $m$ is greater than or equal to 4, the estimation error vector $e_{ei}$ can be approximately reconstructed by the visual measurement error $f_{ei} := f_i - \bar{f}_i$ as
\begin{eqnarray} \label{eq:eei}
e_{ei} = J_i^{\dagger}(\bar{g}_i) f_{ei}.
\end{eqnarray}

In case of a single camera, \cite{TCST07} 
presents an input $u_{ei} = k_e e_{ei}$ based on passivity of the 
estimation error system from $u_{ei}$ to $-e_{ei}$ and the resulting 
estimation mechanism (\ref{eq:EsRRBM}), (\ref{eq:eei}) and $u_{ei} = k_e e_{ei}$ is called \emph{visual motion observer}.
Then, the authors prove the estimate $\bar{g}_i$ converges to the actual relative pose $g_{io_i}$ if $V^b_{wo_i} = 0$.

\cite{CDC11} extended the results in \cite{TCST07}
to the networked vision systems, where
the following input to the model (\ref{eq:EsRRBM})
was proposed.
\begin{eqnarray} \label{eq:input_f}
\!\!\!\!\! u_{ei} = k_e e_{ei} + k_s \sum_{j \in \N_i}E_R(\bar{g}_i^{-1}\bar{g}_{i,j}),\ k_e > 0,\ k_s > 0
\end{eqnarray}
with $\bar{g}_{i,j} := g_{ij}\bar{g}_{j}$.
The input consists of both visual feedback term $k_e e_{ei}$
and mutual feedback term $k_s \sum_{j \in \N_i}E_R(\bar{g}_i^{-1}\bar{g}_{i,j})$
inspired by pose synchronization in \cite{TAC12}.
The resulting networked estimation mechanism (\ref{eq:EsRRBM}), (\ref{eq:eei})
and (\ref{eq:input_f}) is named \emph{networked visual motion observer}.
Then, the paper analyzed the averaging performance
attained by the proposed mechanism.

\subsection{Networked Visual Motion Observer under Imperfect Visibility}

\begin{figure}
\begin{center}
\includegraphics[width=8cm]{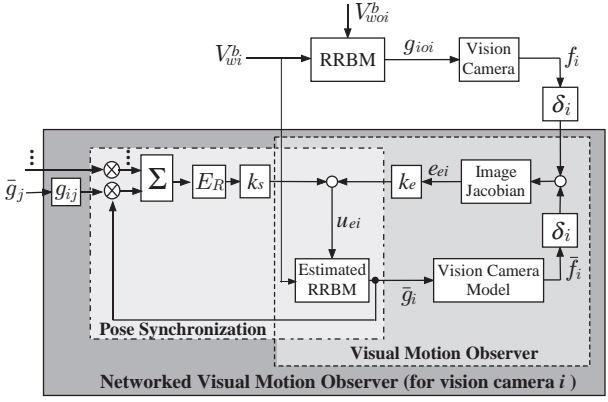}
\caption{Networked visual motion observer}
\label{fig:ce}
\end{center}
\end{figure}

In the presence of the cameras  not capturing the target,
$i \in \V_f^-(t)$ cannot implement the visual feedback term $k_ee_{ei}$
in (\ref{eq:input_f}).
%
We thus employ the following input 
instead of (\ref{eq:input_f}).
\begin{eqnarray} 
u_{ei} = \delta_i(t) k_e e_{ei} +
k_s \sum_{j \in \N_i}E_R(\bar{g}^{-1}_i\bar{g}_{i,j}),
\label{eq:input_f-}
\end{eqnarray}
where $\delta_i(t) = 1$ if $i \in \V_f(t)$ and
$\delta_i(t) = 0$ otherwise.
The total estimation mechanism is formulated as (\ref{eq:EsRRBM}), (\ref{eq:eei}) and the inputs (\ref{eq:input_f-})
whose block diagram with respect to camera $i$ is 
illustrated in Fig. \ref{fig:ce}.

The input (\ref{eq:input_f-}) for $i \in \V_f^-(t)$
is the gradient decent algorithm on $SE(3)$ of the local objective function
$\sum_{j\in \N_i}\psi(\bar{g}_i^{-1}\bar{g}_{i,j})$ (\cite{AbMaSe:08}),
which means each camera in $\V_f^-(t)$ aims at
leading its estimate $\bar{g}_i$ to its neighbors' estimates 
$\{\bar{g}_{i,j}\}_{j\in \N_i}$.
On the other hand, the input for $i \in \V_f(t)$ 
aims at leading $\bar{g}_i$ to both of object pose $g_{io_i}$
and neighbors' estimates.
Meanwhile, the global objective is given by (\ref{eq:g_ave}),
which differs from the local objective functions.
Thus, the closeness between the estimates 
and the global objective minimizer $g^*_i$
is not clear.

In the next section, we thus clarify the averaging performance.
Although it is conjectured from its structure
and demonstrated through simulation 
(\url{http://www.fl.ctrl.titech.ac.jp/researches/movie_new/sim/sw_coopes
t.wmv}) that 
the present mechanism works for a moving object,
we will derive a theoretical result under the 
assumption that the target object is static ($V^b_{wo_i} = 0$).
The main reason to use this assumption is
to assure time invariance of $\V_f(t)$.
Indeed, in case of the time varying $\V_f(t)$, the global objective
itself changes in time and it is necessary to find a metric evaluating 
the performance in order to conduct theoretical analysis, which is left 
as a future work of this paper.


\section{Averaging Performance}
In this section, we derive ultimate estimation accuracy of the average $g^*_i$ achieved by the presented mechanism
assuming that the object is static ($V^b_{wo_i} = 0\ {\forall i}\in \V_f$).
Throughout this section, we use the following assumption.
\begin{assum} \label{assum:static}
 \mbox{ }
\begin{description}
	\item[(i)] The number of elements of $\V_f$ is greater than or 
		   equal to $2$ ($|\V_f| \geq 2$) and there exists a pair $(i,j) \in \V_f \times \V_f$ such that $p_{wo_i} \neq p_{wo_j}$ and $e^{\hat{\xi}\theta_{wo_i}} \neq e^{\hat{\xi}\theta_{wo_j}}$. 
	\item[(ii)] $e^{-\hat{\xi}\theta^*_i}e^{\hat{\xi}\theta_{io_i}} > 0$ for all $i \in \V_f$. 
\end{description}
\end{assum}
The item (i) is assumed just to avoid a meaningless problem such that 
all the poses in $\{g_{wo_i}\}_{i \in \V_f}$ are equal under which
it is straightforward to prove convergence of the estimates to the common pose
by using the techniques presented by \cite{TAC12}.
The detailed discussions on validity of the assumption (ii)
is shown in \cite{TAC13} but it is in general satisfied 
in the scenario of the beginning of Section \ref{sec:problem}.


\subsection{Definition of Approximate Averaging}

In this subsection, we introduce a notion of approximate averaging
similarly to \cite{TAC13}.
For this purpose, we define parameters
\[
 \rho_p := \sum_{i \in {\cal V}_f} \| p_{io_i} - p^*_i \|^2,\
\rho_R := \sum_{i \in {\cal V}_f}\phi(e^{-\hat{\xi}\theta^*_i} e^{\hat{\xi}\theta_{io_i}})
\]
and the following sets for any positive parameter $\varepsilon$.
\begin{eqnarray}
\!\!\!\!\!\Omega_p(\varepsilon) &:=& \Big\{ (\bar{p}_i)_{i \in {\cal V}} \Big| 
\frac{1}{n} \sum_{i \in {\cal V}} \| \bar{p}_i - p^*_i \|^2 <  \varepsilon
\frac{1}{|\V_f|}\rho_p \Big\} \nonumber\\
\!\!\!\!\!\Omega_R(\varepsilon) &:=& \Big\{ (e^{\hat{\bar{\xi}}\bar{\theta}_i})_{i \in {\cal V}} \Big| 
\frac{1}{n} \sum_{i \in {\cal V}} \phi(e^{-\hat{\xi}\theta^*_i}e^{\hat{\bar{\xi}}\bar{\theta}_i}) <  \varepsilon \frac{1}{|\V_f|}\rho_R \Big\} \nonumber
\end{eqnarray}

Let us define $\varepsilon$-level averaging performance to be met by the estimates $\bar{g}_i = (\bar{p}_i, e^{\hat{\bar{\xi}}\bar{\theta}_i})$.

\begin{defn} \label{defn:averaging}
Given target poses $(g_{io_i})_{i \in \V_f}$ and $\varepsilon > 0$, the position estimates $(\bar{p}_i)_{i \in \V}$ and orientation estimates $(e^{\hat{\bar{\xi}}\bar{\theta}_i})_{i \in \V}$
are respectively said to achieve $\varepsilon$-level 
averaging performance, if there exists a finite $T$ such that 
\[
 (\bar{p}_i(t))_{i \in \V} \in \Omega_p(\varepsilon) 
\mbox{ and } (e^{\hat{\bar{\xi}}\bar{\theta}_i}(t))_{i \in \V} \in 
 \Omega_R(\varepsilon)\ \forall t \geq T.
\]
\end{defn}

In case of $\V_f = \V$, $\rho_p$ and $\rho_R$ indicate 
average estimation accuracy in the absence of the mutual feedback 
term of $u_{ei}$ in (\ref{eq:input_f-}) since the visual motion observer correctly estimates the static object pose $g_{io_i}$.
In the case, the parameter $\varepsilon$ is an indicator of improvement of average estimation accuracy by inserting the mutual feedback term.

\subsection{Averaging Performance Analysis}

In this subsection, we state the main result of this paper.
For this purpose, we first define a value
\[
\phi_m := \max_{i \in \V_f} 
\phi(e^{-\hat{\xi}\theta^*_i}e^{\hat{\xi}\theta_{io_i}})
\]
and a parameter $\zeta > 0$ strictly greater than $\phi_m$.
Then, we have the following lemma.
\begin{lem} \label{lem:l1}
Suppose that\\ the targets are static ($V_{wo_i}^b = 0$ $\forall i \in \V_f$) 
 and the estimates $(\bar{g}_i)_{i \in \V}$ are updated according to 
(\ref{eq:EsRRBM}) and (\ref{eq:input_f-}).
Then, under Assumptions 1 and 2 and $e^{-\hat{\bar{\xi}}\bar{\theta}_i}e^{\hat{\xi}\theta^*_i} > 0$ $\forall t \geq 0$, there exists a finite $\tau$ such that $\phi(e^{-\hat{\xi}\theta^*_i}e^{\hat{\bar{\xi}}\bar{\theta}_i}) \leq \zeta$ $\forall t \geq \tau$, $i \in \V$.
\end{lem}
\begin{pf}
See Appendix \ref{sec:pf_lem:l1}
\end{pf}
The proof of Lemma \ref{lem:l1} 
means that the set
\begin{eqnarray}
{\mathcal S} = \{(e^{\hat{\bar{\xi}}\bar{\theta}_i})_{i \in \V} | e^{-\hat{\bar{\xi}}\bar{\theta}_i}e^{\hat{\xi}\theta^*_i} > 0 \ \forall i \in \V \}
\nonumber
\end{eqnarray}
is positively invariant for (\ref{eq:EsRRBM}) with (\ref{eq:input_f-}).


We are now ready to state the main result of this section.
\begin{thm} \label{thm:eps-level}
Suppose the targets are static ($V_{wo_i}^b = 0$ $\forall i \in \V_f$) and the estimates $(\bar{g}_i)_{i \in \V}$ are updated according to (\ref{eq:EsRRBM}) and (\ref{eq:input_f-}).
Then, under Assumptions 1 and 2 and $\beta := 1 - \sqrt{2\zeta} > 0$, if 
 the initial estimates satisfy 
 $(e^{\hat{\bar{\xi}}\bar{\theta}_i}(0))_{i \in \V} \in {\mathcal S}$, 
 for any $\epsilon \in (0,1)$, there exists a sufficiently small $k=k_e/k_s$
such that the position estimates $(\bar{p}_i)_{i \in \V}$ achieve 
 $\epsilon$-level averaging performance and the orientation estimates $(e^{\hat{\bar{\xi}}\bar{\theta}_i})_{i \in \V}$ achieve $\varepsilon_R$-level averaging performance with
$\varepsilon_R = 1 - (1 - \epsilon)\beta$. 
\end{thm}
\begin{pf}
See Appendix \ref{app:1}.
\end{pf}
Theorem \ref{thm:eps-level} says that choosing the gains $k_e$ and $k_s$
such that $k = k_e/k_s$ is sufficiently small leads to a good
averaging performance.
The conclusion is the same as \cite{CDC11,TAC13} and hence
the contribution of the theorem is to prove the statement is also
valid even in the presence of the cameras not viewing the target.
We also see an essential difference between the position and orientation 
estimates that the averaging performance on positions
can be arbitrarily improved by choosing a sufficiently small $k$ but 
an offset associated with $\beta < 1$ occurs for the orientation estimates.

\begin{rm}
The energy function $U_R$ in (\ref{eq:energy}), which allows us to prove Theorem \ref{thm:eps-level},
is defined by the sum of individual error between the 
average and the estimate. 
The selection of this function is inspired by \cite{chopra}.
\end{rm}


\section{Verification through simulation}
\begin{figure}
\begin{center}
\includegraphics[width=7cm]{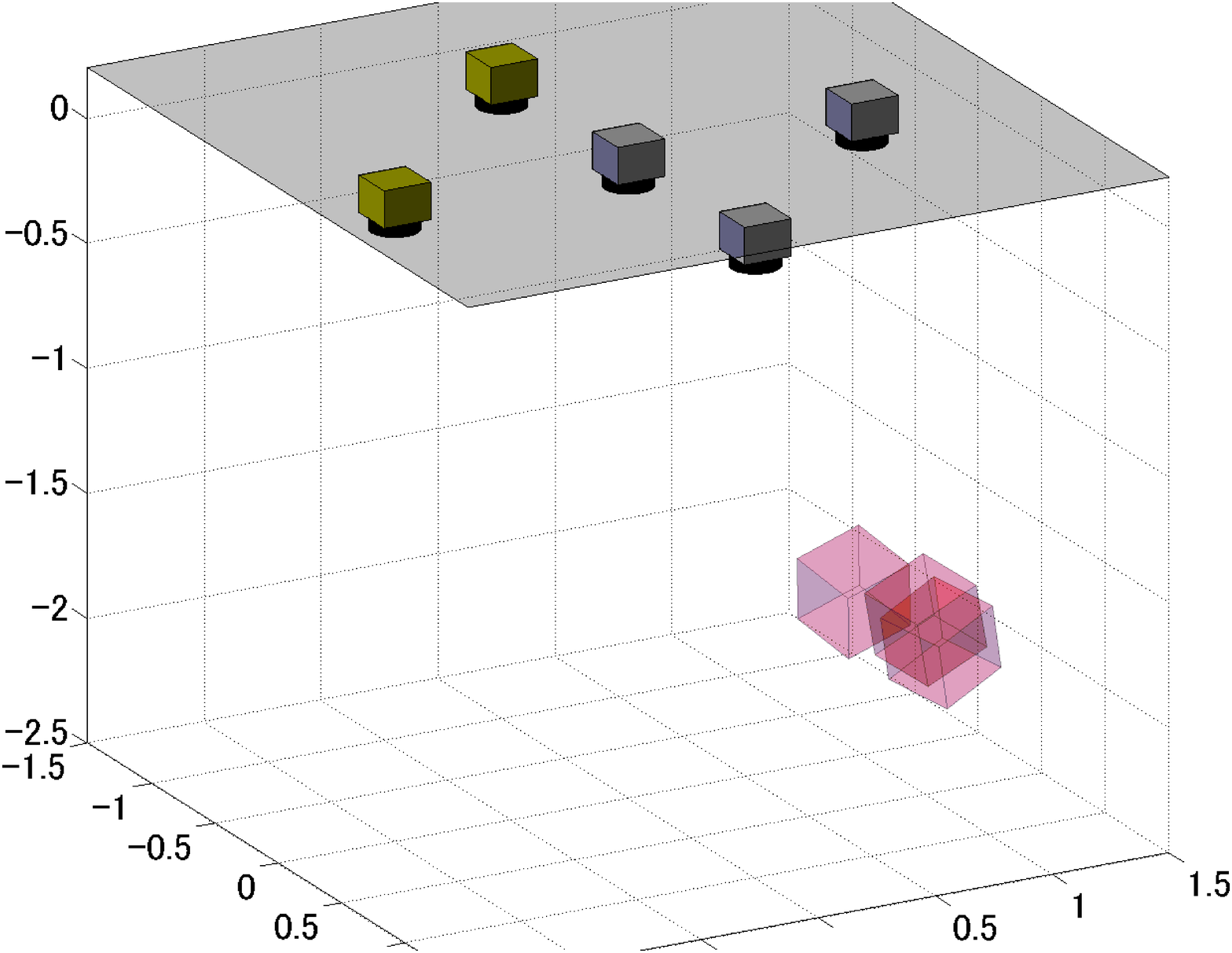}    
\caption{Overview of simulation} 
\label{fig:overview}
\begin{minipage}{4.2cm}
\includegraphics[width=4.2cm]{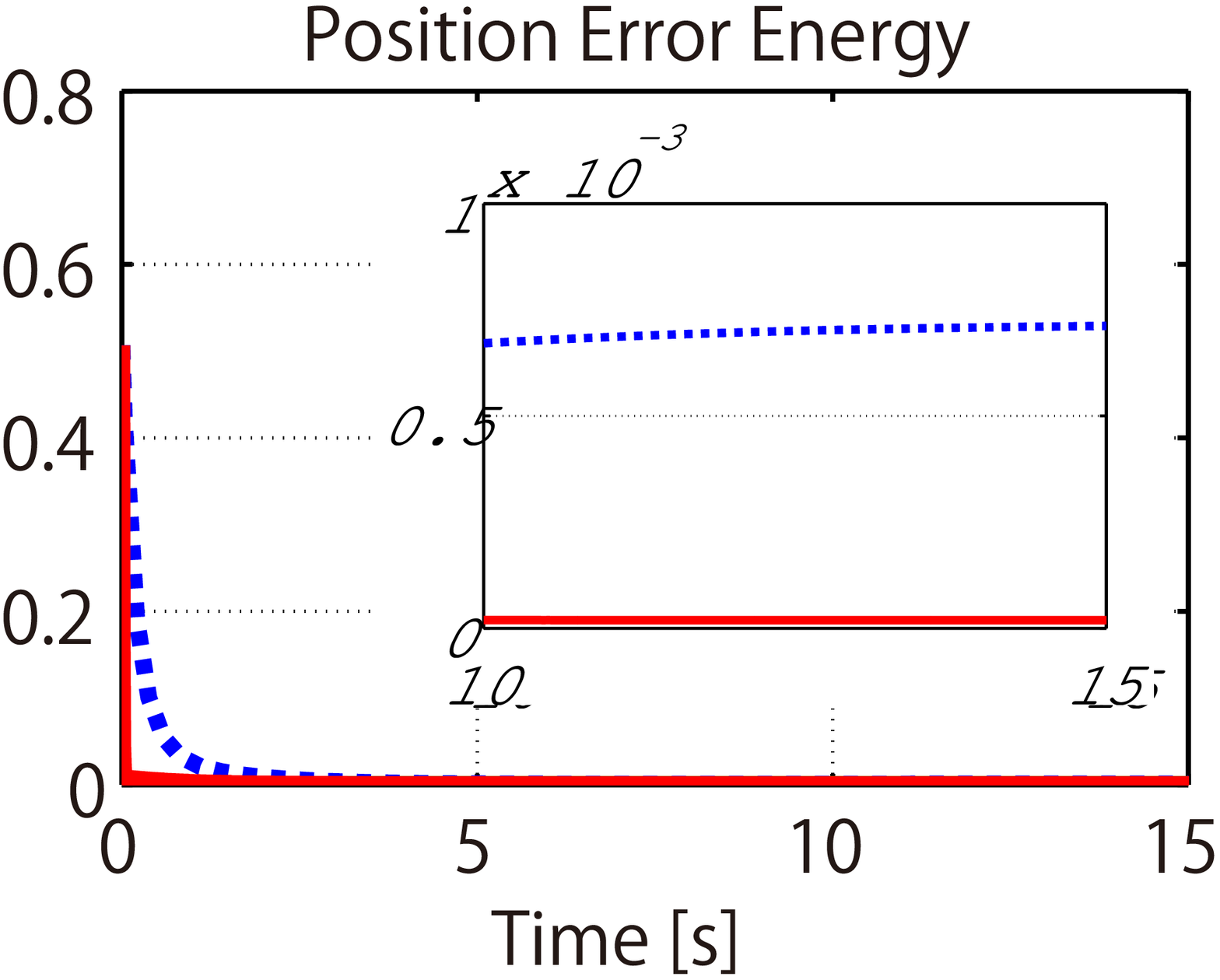}    
\end{minipage}
\begin{minipage}{4.2cm}
\includegraphics[width=4.2cm]{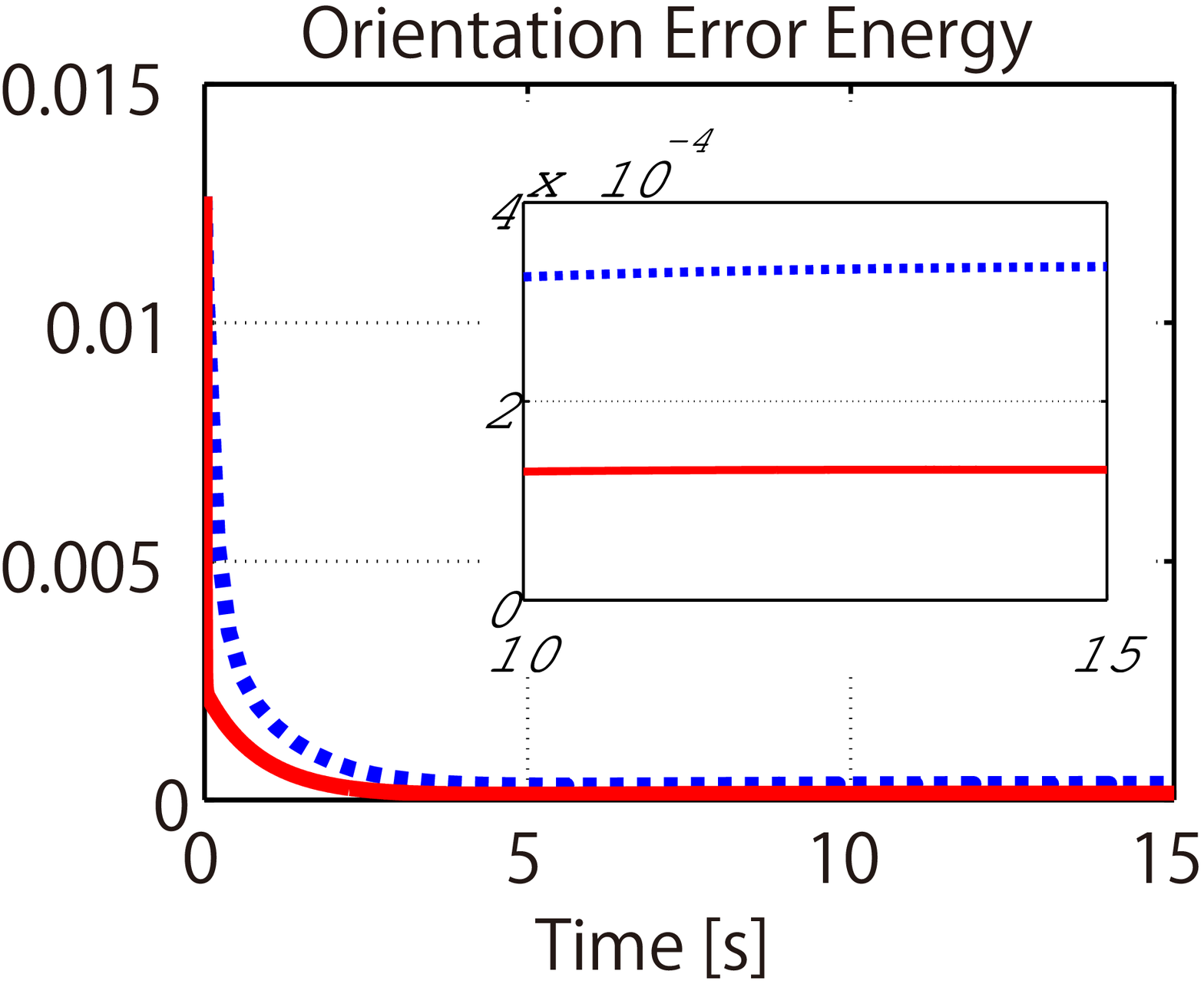}    
\end{minipage}
\caption{Time responses of estimation error energies 
for $k_s = 1$ and $k_s = 50$} 
\label{fig:energy}
\end{center}
\end{figure}

We finally demonstrate the effectiveness of the present algorithm through simulation.
Here, we consider five pin-hole type cameras with focal length 
$0.03\,\mathrm{m}$ connected by
the communication graph with
${\mathcal E} = \{(1,2),(1,3),(1,5),$ $(2,3),(3,4),(4,5)\}$.
We identify the frame of camera 1 with the world frame and let $p_{w2} = [
\begin{array}{ccc}
	1 & 0 & 0 \\
\end{array}
]^T, \ p_{w3} = [
\begin{array}{ccc}
	0 & 1 & 0 \\
\end{array}
]^T, \ p_{w4} = [
\begin{array}{ccc}
	-1 & 0 & 0 \\
\end{array}
]^T, \ p_{w5} = [
\begin{array}{ccc}
	0 & -1 & 0 \\
\end{array}
]^T$ and $e^{\hat{\xi}\theta_{wi}} = I_3, \forall i \in \{2,3,4,5\}$.
Let only cameras $i=1,2,3$ (gray boxes in Fig. \ref{fig:overview}) 
capture the target, i.e. $\V_f = \{1,2,3\}$.

We set the configurations of target objects as $p_{wo_1} = [
\begin{array}{ccc}
	0.55 & 1.00 & -1.91 \\
\end{array}
]^T, \ p_{wo_2} = [
\begin{array}{ccc}
	0.30 & 0.80 & -1.84 \\
\end{array}
]^T, \ p_{wo_3} = [
\begin{array}{ccc}
	0.56 & 1.05 & -2.00 \\
\end{array}
]^T, \ \xi\theta_{wo_1} = [
\begin{array}{ccc}
	0.30 & 0.19 & 0.21 \\
\end{array}
]^T, \ \xi\theta_{wo_2} = [
\begin{array}{ccc}
	0.21 & 0.30 & 0.19 \\
\end{array}
]^T, \ \xi\theta_{wo_3} = [
\begin{array}{ccc}
	0.29 & 0.20 & 0.31 \\
\end{array}
]^T$.
The red boxes in Fig. \ref{fig:overview} represent the initial configuration of target objects and yellow boxes represent the cameras $\V_f^- = \{4, 5\}$.
Then, the average $g^* = (p^*, e^{\hat{\xi}\theta^*})$ is given by
$p^* = [
\begin{array}{ccc}
	0.47 & 0.95 & -1.92 \\
\end{array}
]^T,\ \xi\theta^* = [
\begin{array}{ccc}
	0.27 & 0.23 & 0.24 \\
\end{array}
]^T$.

\begin{figure}
\begin{center}
\includegraphics[width=8.4cm]{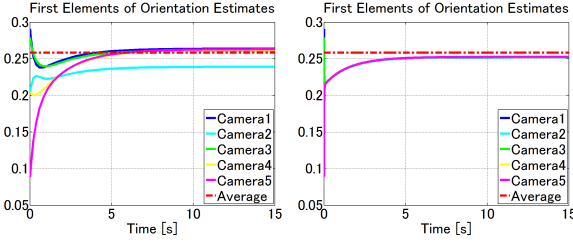}    
\caption{Time responses of the first element of $\bar{\xi} \sin(\bar{\theta}_{w,i})$ \newline $i \in 1, \cdots, 5$ (Left: $k_s = 1$, Right $k_s = 50$)} 
\label{fig:xix}
\end{center}
\end{figure}

We run simulations with two different gains $k_e = 1,\ k_s = 1\ (k = 
1)$ and $k_e = 1,\ k_s = 50\ (k = 0.02)$ from the initial condition
$\bar{p}_i(0) = [\begin{array}{ccc}
	0 & 0 & 1 \\
\end{array}
]^T$ and $e^{\hat{\bar{\xi}}\bar{\theta}_i}(0) = I_3\ {\forall i}$.
Fig. \ref{fig:energy} shows the time responses of the position estimation 
error energy
\[
 U_p = \frac{1}{2}\sum_{i\in \V} \|\bar{p}_{i} - p_i^*\|^2
\]
and orientation estimation error energy $U_R$ defined in (\ref{eq:energy}),
where the red solid curves illustrate the result for $k_s = 50$
and the blue dashed curves that for $k_s = 1$.
We see from both figures that the energies for the larger mutual feedback 
gain $k_s = 50$ are smaller than those for  $k_s=1$, 
which implies that a large $k_s$ and hence a small $k$ achieves
a good averaging performance as indicated by
Theorem \ref{thm:eps-level}.
Fig. \ref{fig:xix} illustrates the time responses of the
first element of orientation estimates $\bar{\xi}\sin(\bar{\theta}_{w,i})
(e^{\bar{\xi}\bar{\theta}_{w,i}} = 
e^{\hat{\xi}\theta_{wi}}e^{\bar{\xi}\bar{\theta}_{i}})$ 
of all cameras produced by the networked visual motion observer, 
where the red dash-dotted line represents the average.
We also see from the figure that, while the estimate of camera 2 
for $k_s=1$ is far from the average, all the estimates for $k_s = 50$
approaches to it.
However, we also confirm that an offset still occurs even in case
of $k_s = 50$ as indicated in Theorem \ref{thm:eps-level}.


\section{Conclusion}

In this paper, we have investigated a vision-based cooperative estimation
problem of a 3D target object pose for visual sensor networks.
In particular, we have extended the 
networked visual motion observer presented by \cite{CDC11}
so that it works even in the presence of cameras not viewing the target
due to the limited view angles and obstructions.
Then, we have analyzed the averaging performance attained by the present mechanism.
Finally, we have demonstrated the effectiveness of the present algorithm
through simulation.



\appendix


\section{Proof of Lemma \ref{lem:l1}}    
\label{sec:pf_lem:l1}

In the proof, we use the following lemma.
\begin{lem}[\cite{TAC12}] \label{lem:sym}
For any matrices $R_1,$ $R_2, R_3 \in SO(3)$, the inequality 
\begin{eqnarray} \label{eq:sym}
\frac{1}{2} \tr (R_1^TR_2 &-& R_1^TR_3 R_2^T R_3) \geq \phi(R_1^T R_3)  - \phi(R_1^T R_2) \nonumber \\
&+& \lambda_{min} (\sym(R_1^TR_3)) \phi(R_3^TR_2) 
\end{eqnarray}
holds, where $\sym(M) := \frac{1}{2}(M + M^T)$ and $\lambda_{min}(M)$ is the minimal eigenvalue of matrix $M$.
\end{lem}

Extracting the time evolution of the orientation estimates 
from (\ref{eq:EsRRBM}) with (\ref{eq:input_f-}) 
and transforming their coordinates from $\Sigma_i$
to $\Sigma_w$ as 
$e^{\hat{\bar{\xi}}\bar{\theta}_{w,i}} = 
e^{\hat{\xi}\theta_{wi}}e^{\hat{\bar{\xi}}\bar{\theta}_{i}}$ 
yields
\begin{eqnarray} 
\dot{e}^{\hat{\bar{\xi}}\bar{\theta}_{w,i}} &=& e^{\hat{\bar{\xi}}\bar{\theta}_{w,i}}\hat{\omega}_{uei}, 
\label{eq:es_ori_ff}\\
\omega_{uei} &=& \delta_i k_e e_R(e^{-\hat{\bar{\xi}}\bar{\theta}_{w,i}}e^{\hat{\xi}\theta_{wo_i}}) 
\nonumber\\
&&\hspace{1cm}+ k_s \sum_{j \in \N_i} e_R(e^{-\hat{\bar{\xi}}\bar{\theta}_{w,i}}e^{\hat{\bar{\xi}}\bar{\theta}_{w,j}}), \label{eq:es_ori_f}
\end{eqnarray}
which is independent of evolution of the position estimates.

Let us now consider the energy function
\begin{eqnarray} \label{eq:U}
U &:=& \phi(e^{-\hat{\xi}\theta^*_l}e^{\hat{\bar{\xi}}\bar{\theta}_l}) = \phi(e^{-\hat{\xi}\theta^*}e^{\hat{\bar{\xi}}\bar{\theta}_{w,l}}),
\end{eqnarray}
similarly to Lemma 1 in \cite{TAC13},
where $l(t) := \arg \max_{i \in {\cal V}} \phi(e^{-\hat{\xi}\theta^*_i}e^{\hat{\bar{\xi}}\bar{\theta}_i}(t))$.
The time derivative of $U$ along with the trajectories 
of (\ref{eq:es_ori_ff}) with (\ref{eq:es_ori_f}) is given as
\begin{eqnarray} 
\dot{U} &=& 2 e_R^T(e^{-\hat{\xi}\theta^*}e^{\hat{\bar{\xi}}\bar{\theta}_{w,l}})\omega_{uel} 
\nonumber\\
&=& -\tr\left(\sk(e^{-\hat{\xi}\theta^*}e^{\hat{\bar{\xi}}\bar{\theta}_{w,l}})\hat{\omega}_{uel}\right), \label{eq:dotU}
\end{eqnarray}
where we use the relation $a^Tb = - \frac{1}{2} \tr (\hat{a}\hat{b})$.
Substituting  (\ref{eq:es_ori_f}) into (\ref{eq:dotU}) yields
\begin{eqnarray} \label{eq:dotU1}
&\dot{U}&= -\frac{1}{2} \tr \Bigl\{ \delta_{l(t)}k_e (e^{-\hat{\xi}\theta^*}e^{\hat{\xi}\theta_{wo_l}} 
	- e^{-\hat{\xi}\theta^*}e^{\hat{\xi}\theta_{wo_l}}e^{-\hat{\bar{\xi}}\bar{\theta}_{w,l}}e^{\hat{\xi}\theta_{wo_l}}) \nonumber \\
&+& k_s \sum_{j \in {\cal N}_{l}} (e^{-\hat{\xi}\theta^*}e^{\hat{\bar{\xi}}\bar{\theta}_{w,j}} 
	- e^{-\hat{\xi}\theta^*}e^{\hat{\bar{\xi}}\bar{\theta}_{w,l}}e^{\hat{\bar{\xi}}\bar{\theta}_{w,j}}e^{\hat{\bar{\xi}}\bar{\theta}_{w,l}}) \Bigr\}.
\end{eqnarray}
From Lemma \ref{lem:sym}, (\ref{eq:dotU1}) is rewritten as 
\[
 \dot{U} \leq - \delta_{l(t)}k_e A_l - k_s B_l,
\]
where
\begin{eqnarray}
A_i &:=& \phi(e^{-\hat{\xi}\theta^*}e^{\hat{\bar{\xi}}\bar{\theta}_{w,i}}) - \phi(e^{-\hat{\xi}\theta^*}e^{\hat{\xi}\theta_{wo_i}}) + \sigma_i \phi(e^{-\hat{\bar{\xi}}\bar{\theta}_{w,i}} e^{\hat{\xi}\theta_{wo_i}}), \nonumber \\  
\sigma_i &:=& \lambda_{min}(\sym (e^{-\hat{\xi}\theta^*}e^{\hat{\bar{\xi}}\bar{\theta}_{w,i}})), \nonumber \\
B_i &:=& \sum_{j \in {\cal N}_{i}}\Big(\phi(e^{-\hat{\xi}\theta^*}e^{\hat{\bar{\xi}}\bar{\theta}_{w,i}}) - \phi(e^{-\hat{\xi}\theta^*}e^{\hat{\bar{\xi}}\bar{\theta}_{w,j}}) \nonumber \\
&& \indent \indent \indent \indent \hspace{4cm} + \sigma_i \phi(e^{-\hat{\bar{\xi}}\bar{\theta}_{w,i}} e^{\hat{\bar{\xi}}\bar{\theta}_{w,j}})\Big). \nonumber 
\end{eqnarray}
The inequality $\phi(e^{-\hat{\xi}\theta^*}e^{\hat{\bar{\xi}}\bar{\theta}_{w,l}}) \geq \phi(e^{-\hat{\xi}\theta^*}e^{\hat{\bar{\xi}}\bar{\theta}_{w,j}})$ $\forall j \in \V$ holds from the definition of the index $l$, and hence we obtain $B_l \geq \sum_{j \in {\cal N}_{l}} \sigma_l \phi(e^{-\hat{\bar{\xi}}\bar{\theta}_{w,l}} e^{\hat{\bar{\xi}}\bar{\theta}_{w,j}})$.
Thus, the inequality
\begin{eqnarray*}
\dot{U} &\leq& -\Bigl(\delta_{l(t)}k_e \phi(e^{-\hat{\xi}\theta^*}e^{\hat{\bar{\xi}}\bar{\theta}_{w,l}}) - \delta_{l(t)}k_e \phi(e^{-\hat{\xi}\theta^*}e^{\hat{\xi}\theta_{wo_l}}) \nonumber \\
&+& \sigma_l (\delta_{l(t)}k_e \phi(e^{-\hat{\bar{\xi}}\bar{\theta}_{w,l}} e^{\hat{\xi}\theta_{wo_l}}) 
+ k_s \sum_{j \in {\cal N}_{l}} \phi(e^{-\hat{\bar{\xi}}\bar{\theta}_{w,l}} e^{\hat{\bar{\xi}}\bar{\theta}_{w,j}}))\Bigr)
\label{eq:dotU2}
\end{eqnarray*}
is true.
From Assumption \ref{assum:static}, we have $\sigma_l > 0$ and hence
\begin{eqnarray} \label{eq:dotU3}
\dot{U} &\leq& -\delta_{l(t)}k_e (\phi(e^{-\hat{\xi}\theta^*}e^{\hat{\bar{\xi}}\bar{\theta}_{w,l}}) - \phi(e^{-\hat{\xi}\theta^*}e^{\hat{\xi}\theta_{wo_l}})).
\end{eqnarray}

Suppose now that $l(t) \in \V_f$.
Then, if 
$\phi(e^{-\hat{\xi}\theta^*}e^{\hat{\bar{\xi}}\bar{\theta}_{w,l}}) > 
\zeta$, $\dot{U} < 0$ is true from the definition of $\zeta$.
On the other hand, in case of $l(t) \in \V_f^-$, we also have $\dot{U} \leq 0$.
Namely, the function $U$ never increases as long as an estimate
$e^{\hat{\bar{\xi}}\bar{\theta}_{w,i}}$ satisfies
$\phi (e^{-\hat{\xi}\theta^*}e^{\hat{\bar{\xi}}\bar{\theta}_{w,i}})
> \zeta$.
This implies 
that once the estimates $(e^{\hat{\bar{\xi}}\bar{\theta}_{w,i}})_{i\in 
\V}$ enter 
\[
{\mathcal S}_{\zeta} := \{(e^{\hat{\bar{\xi}}\bar{\theta}_{w,i}})_{i\in 
\V}|\ \phi (e^{-\hat{\xi}\theta^*}e^{\hat{\bar{\xi}}\bar{\theta}_{w,i}})
\leq \zeta\ 
{\forall i}\in \V\}
\]
at a time,
$(e^{\hat{\bar{\xi}}\bar{\theta}_{w,i}})_{i\in \V}$ stays
in the set ${\mathcal S}_{\zeta}$ for all subsequent time.

Let us now employ another energy function
\begin{eqnarray}
V &:=& \sum_{i \in \Lambda(t)}(\phi(e^{-\hat{\xi}\theta^*}e^{\hat{\bar{\xi}}\bar{\theta}_{w,i}}) - \zeta),
\label{eq:V} \\
\Lambda(t) &:=& \{{i \in {\cal V}} | \phi(e^{-\hat{\xi}\theta^*}e^{\hat{\bar{\xi}}\bar{\theta}_{w,i}}) \geq \zeta \}. \label{eq:Lambda} 
\end{eqnarray}
The function $V \geq 0$ is continuous but it may not be differentiable 
on the region where an estimate $e^{\hat{\bar{\xi}}\bar{\theta}_{w,i}}$
satisfies $\phi(e^{-\hat{\xi}\theta^*}e^{\hat{\bar{\xi}}\bar{\theta}_{w,i}})= \zeta$.
Except for the region, the time derivative of $V$ along with the trajectories of 
(\ref{eq:es_ori_ff}) is given by 
\begin{equation}
 \dot{V} \leq -k_e \sum_{i \in \Lambda(t)\cap \V_f} A_i -k_s \sum_{i \in \Lambda(t)} B_i.
\label{eq_hata1}
\end{equation}
We first consider 
\begin{equation}
 \sum_{i \in \Lambda(t)} \sum_{j \in {\cal N}_{i}} \left(
\phi(e^{-\hat{\xi}\theta^*}e^{\hat{\bar{\xi}}\bar{\theta}_{w,i}}) - 
\phi(e^{-\hat{\xi}\theta^*}e^{\hat{\bar{\xi}}\bar{\theta}_{w,j}})\right)
\label{eq_hata6}
\end{equation}
in the second term of .
In case of $j \notin \Lambda(t)$, we have 
\[
 \phi(e^{-\hat{\xi}\theta^*}e^{\hat{\bar{\xi}}\bar{\theta}_{w,i}}) - \phi(e^{-\hat{\xi}\theta^*}e^{\hat{\bar{\xi}}\bar{\theta}_{w,j}}) > 0.
\]
Otherwise ($j \in \Lambda(t)$), the term 
\[
 \phi(e^{-\hat{\xi}\theta^*}e^{\hat{\bar{\xi}}\bar{\theta}_{w,j}}) - 
\phi(e^{-\hat{\xi}\theta^*}e^{\hat{\bar{\xi}}\bar{\theta}_{w,i}})
\]
has to appear in (\ref{eq_hata6})
under Assumption 1 and 
they are canceled.
Thus, the inequality 
\[
\sum_{i \in \Lambda(t)} \sum_{j \in {\cal N}_{i}} \phi(e^{-\hat{\xi}\theta^*}e^{\hat{\bar{\xi}}\bar{\theta}_{w,i}}) 
	- \phi(e^{-\hat{\xi}\theta^*}e^{\hat{\bar{\xi}}\bar{\theta}_{w,j}}) \geq 0 
\] 
holds and hence we obtain
\begin{eqnarray} \label{eq:b_i}
\sum_{i \in \Lambda} B_i \geq \sum_{i \in \Lambda(t)}\sum_{j \in {\cal N}_{i}} \sigma_i \phi(e^{-\hat{\bar{\xi}}\bar{\theta}_{w,i}} e^{\hat{\bar{\xi}}\bar{\theta}_{w,j}}).
\end{eqnarray}
Note that the equality of (\ref{eq:b_i}) can hold only if $ \Lambda (t) = \V$
or $\Lambda(t) = \emptyset$
since, otherwise, there must be a pair of $(i,j)\in {\mathcal E}$ such that
$j \notin \Lambda(t)$ and $i \in \Lambda(t)$ from Assumption 1.

Now, substituting (\ref{eq:b_i}) into (\ref{eq_hata1}) yields
\begin{eqnarray}
&\dot{V}& \leq - k_e \sum_{i \in \Lambda(t)\cap \V_f}\Big(\phi(e^{-\hat{\xi}\theta^*}e^{\hat{\bar{\xi}}\bar{\theta}_{w,i}}) 
- \phi(e^{-\hat{\xi}\theta^*}e^{\hat{\xi}\theta_{wo_i}}) \nonumber \\
&+& \sigma_i \phi(e^{-\hat{\bar{\xi}}\bar{\theta}_{w,i}} e^{\hat{\xi}\theta_{wo_i}}) \Big)
- k_s \sum_{i \in \Lambda(t)} 
\sigma_i\sum_{j \in \N_i} \phi(e^{-\hat{\bar{\xi}}\bar{\theta}_{w,i}} e^{\hat{\bar{\xi}}\bar{\theta}_{w,j}})
. \nonumber
\end{eqnarray}
We see from the inequality and the definition of $\zeta$ that
if there exists $i \in \Lambda(t)\cap \V_f$ then
$\dot{V} < 0$.
In addition, if $\Lambda(t)\cap \V_f = \emptyset$,
then the inequality (\ref{eq:b_i}) strictly holds and hence $\dot{V}< 0$.
Namely, the function $V$ is strictly decreasing 
except for the region where an estimate $e^{\hat{\bar{\xi}}\bar{\theta}_{w,i}}$
satisfies 
\begin{equation}
\phi(e^{-\hat{\xi}\theta^*}e^{\hat{\bar{\xi}}\bar{\theta}_{w,i}})= \zeta.
\label{eq_hata5}
\end{equation}
Since the function $V$ is continuous despite of
the event that an estimate goes across the region,
the function $V$ decreases and
the estimates $(e^{\hat{\bar{\xi}}\bar{\theta}_{w,i}})_{i\in \V}$
enters ${\mathcal S}_{\zeta}$ at least once
as long as the time interval 
\[
 T_{b} := \{t\geq 0|\ \exists i\in \V \mbox{ satisfying (\ref{eq_hata5})}\}
\]
is bounded 
and the number of occurrences of the event (\ref{eq_hata5})
is finite.
As proved above, if all the estimates enter ${\mathcal S}_{\zeta}$ once
they have to stay there for all subsequent time.

Let us now define $V'$
and $\Lambda'(t)$ by just replacing $\zeta$ by $\zeta'= \phi_m + (\zeta-\phi_m)/2$.
Then, all the above discussions hold true.
Notice that every time an estimate goes across the region of
(\ref{eq_hata5}),
the estimate has to spend nonzero finite time in the region where
$\phi(e^{-\hat{\xi}\theta^*}e^{\hat{\bar{\xi}}\bar{\theta}_{w,i}}) \in 
(\zeta', \zeta)$ from continuity of $e^{\hat{\bar{\xi}}\bar{\theta}_{w,i}}$
and $\|\omega_{uei}\|<\infty$.
During the period, the function $V'$ is strictly decreasing.
Namely, if the event happens infinitely often, $V' \rightarrow - \infty$,
which contradicts $V' \geq 0$.
The possibility that $T_{b}$ is unbounded is also excluded in the same way.
This completes the proof.

\section{Proof of Theorem \ref{thm:eps-level}}
\label{app:1}

In this paper, we prove only the orientation part
since it is possible to prove the position part
in the same way.
The evolution of orientation estimates
$(e^{\hat{\bar{\xi}}\bar{\theta}_{w,i}})_{i\in \V}$
is described by (\ref{eq:es_ori_ff}) with (\ref{eq:es_ori_f}).

We first define the energy function
\begin{eqnarray}
U_R &:=& \sum_{i \in \V}\phi(e^{-\hat{\xi}\theta^*_i}e^{\hat{\bar{\xi}}\bar{\theta}_i}) = 
\sum_{i \in \V}\phi(e^{-\hat{\xi}\theta^*}e^{\hat{\bar{\xi}}\bar{\theta}_{w,i}})
\label{eq:energy}
\end{eqnarray}
and the sets 
\begin{eqnarray*}
\hspace{-.2cm}&&{\cal S}_1(\varepsilon) := \Big\{(e^{\hat{\bar{\xi}}\bar{\theta}_{w,i}})_{i \in {\cal V}} \in {\cal S} \Big| \sum_{i \in \V_f}\phi(e^{-\hat{\xi}\theta^*}e^{\hat{\bar{\xi}}\bar{\theta}_{w,i}}) > \varepsilon \rho_R \Big\}, \nonumber \\
\hspace{-.2cm}&&{\cal S}_2 := \Big\{(e^{\hat{\bar{\xi}}\bar{\theta}_{w,i}})_{i \in {\cal V}} \in {\cal S} \Big| 
\sum_{i \in \V}\sum_{j \in \N_i}\phi(e^{-\hat{\bar{\xi}}\bar{\theta}_{w,i}}e^{\hat{\bar{\xi}}\bar{\theta}_{w,j}}) > \frac{k \rho_R}{\beta}  \Big\}. \nonumber
\label{eq:set_R}
\end{eqnarray*}
Then, we first prove the following lemma.
\begin{lem} \label{lem:l2}
Suppose that all the assumptions of Theorem \ref{thm:eps-level}
hold. Then, there exists a sufficiently small $k$
such that the time derivative of $U_R$
along with the trajectories of (\ref{eq:es_ori_ff}) and 
 (\ref{eq:es_ori_f}) satisfies $\dot{U}_R < 0$ at least after the time $\tau$
in the region where
 $(e^{\hat{\bar{\xi}}\bar{\theta}_{w,i}})_{i \in {\cal V}} \in ({\cal 
 S}_1(\varepsilon_R') \cup {\cal S}_2)$ with 
\[
 \varepsilon_R' := 1 - (1-\epsilon)(\sqrt{\beta}-\sqrt{kW})^2.
\]
\end{lem}
\begin{pf}
The time derivative of $U_R$ along the trajectories of 
 (\ref{eq:es_ori_ff}) and (\ref{eq:es_ori_f}) is given by
\begin{eqnarray}
\!\!\!\!\!\!\!\!\!\!\dot{U}_R &=& 2 \sum_{i \in \V} e_R^T(e^{-\hat{\xi}\theta^*}e^{\hat{\bar{\xi}}\bar{\theta}_{w,i}})\omega_{uei}
\nonumber\\
&=& -\tr\left(\sk(e^{-\hat{\xi}\theta^*}e^{\hat{\bar{\xi}}\bar{\theta}_{w,i}})\hat{\omega}_{uei}\right).
\label{eq:dotUR}
\end{eqnarray}
Substituting  (\ref{eq:es_ori_f}) into (\ref{eq:dotUR}) yields
\begin{eqnarray}
\dot{U}_R &=& -k_e\sum_{i \in \V_f}{\tr}(\Phi_1) - k_s\sum_{i \in {\cal V}}{\tr}(\Phi_2),
\label{eq:dotUR1} \\
\Phi_1 &=& \frac{1}{2}\Big(e^{-\hat{\xi}\theta^*}e^{\hat{\bar{\xi}}\bar{\theta}_{w,i}} 
- e^{-\hat{\xi}\theta^*}e^{\hat{\bar{\xi}}\bar{\theta}_{w,i}}e^{-\hat{\xi}\theta_{wo_i}}e^{\hat{\bar{\xi}}\bar{\theta}_{w,i}}\Big), \nonumber \\
\Phi_2 &=& \frac{1}{2}\sum_{j \in {\cal N}_i}\Big(e^{-\hat{\xi}\theta^*} e^{\hat{\bar{\xi}}\bar{\theta}_{w,j}} 
- e^{-\hat{\xi}\theta^*} e^{\hat{\bar{\xi}}\bar{\theta}_{w,i}}e^{-\hat{\bar{\xi}}\bar{\theta}_{w,j}}e^{\hat{\bar{\xi}}\bar{\theta}_{w,i}}\Big). \nonumber 
\end{eqnarray}

We first consider the term $\sum_{i \in \V}\tr(\Phi_2)$.
From Lemma \ref{lem:sym}, the following inequality holds.
\begin{eqnarray} \label{eq:Phi2}
&\sum_{i \in \V}& \tr (\Phi_2) \geq 
\sum_{i \in {\cal V}}\sum_{j \in {\cal N}_i} \Big\{\phi(e^{-\hat{\xi}\theta^*}e^{\hat{\bar{\xi}}\bar{\theta}_{w,i}}) \nonumber \\ 
&& - \phi(e^{-\hat{\xi}\theta^*}e^{\hat{\bar{\xi}}\bar{\theta}_{w,j}})
+ \sigma_i \phi(e^{-\hat{\bar{\xi}}\bar{\theta}_{w,i}} e^{\hat{\bar{\xi}}\bar{\theta}_{w,j}}) \Big\}. 
\end{eqnarray}
Assumption \ref{assum:graph} implies that 
\[
\sum_{i \in \V}\sum_{j \in \N_i} \phi(e^{-\hat{\xi}\theta^*}e^{\hat{\bar{\xi}}\bar{\theta}_{w,i}}) 
- \phi(e^{-\hat{\xi}\theta^*}e^{\hat{\bar{\xi}}\bar{\theta}_{w,j}}) = 0 
\]
(\cite{TAC12}).
From Lemma \ref{lem:l1}, (\ref{eq:Phi2}) is rewritten as
\begin{eqnarray} \label{eq:Phi2_1}
\sum_{i \in \V} \tr (\Phi_2) &\geq& \sum_{i \in \V}\sum_{j \in \N_i} \beta \phi(e^{-\hat{\bar{\xi}}\bar{\theta}_{w,i}} e^{\hat{\bar{\xi}}\bar{\theta}_{w,j}})
\end{eqnarray}
at least after time $\tau$ similarly to \cite{TAC13}.

We next consider the term $\sum_{i \in \V_f}\tr(\Phi_1)$ in (\ref{eq:dotUR1}).
In the same way as \cite{TAC13}, we can prove
\begin{eqnarray}
\hspace{-.6cm}&& \sum_{i \in {\cal V}_f} \tr (\Phi_1) \geq
\sum_{i \in {\V}_f}\Big\{\phi(e^{-\hat{\xi}\theta^*}e^{\hat{\bar{\xi}}\bar{\theta}_{w,i}}) - \phi(e^{-\hat{\xi}\theta^*} e^{\hat{\xi}\theta_{wo_i}}) \nonumber \\
\hspace{-.6cm}&&\hspace{4cm}+  \beta \phi(e^{-\hat{\bar{\xi}}\bar{\theta}_{w,i}} e^{\hat{\xi}\theta_{wo_i}})\Big\}.
\nonumber\\
\hspace{-.6cm}&&= -\rho_R + \sum_{i \in {\V}_f}\Big\{ \phi(e^{-\hat{\xi}\theta^*}e^{\hat{\bar{\xi}}\bar{\theta}_{w,i}}) 
+  \beta \phi(e^{-\hat{\bar{\xi}}\bar{\theta}_{w,i}} e^{\hat{\xi}\theta_{wo_i}})
\Big\}
\label{eq:Phi1}
\end{eqnarray}
at least after time $\tau$.
Substituting (\ref{eq:Phi1}) and (\ref{eq:Phi2_1}) into (\ref{eq:dotUR1}) yields
\begin{eqnarray}
\dot{U}_R &\leq& - k_e \sum_{i \in \V_f}\Big\{\phi(e^{-\hat{\xi}\theta^*}e^{\hat{\bar{\xi}}\bar{\theta}_{w,i}}) + \beta \phi(e^{-\hat{\bar{\xi}}\bar{\theta}_{w,i}} e^{\hat{\xi}\theta_{wo_i}})) \Big\} \nonumber \\
 && + k_e \rho_R - k_s \sum_{i \in \V}\sum_{j \in \N_i} \beta \phi(e^{-\hat{\bar{\xi}}\bar{\theta}_{w,i}} e^{\hat{\bar{\xi}}\bar{\theta}_{w,j}}).
\label{eq:dotUR2}
\end{eqnarray}
From (\ref{eq:dotUR2}), $\beta>0$ and the definitions of the sets ${\mathcal S}_1(\varepsilon)$ 
 and ${\mathcal S}_2$,
we have $\dot{U}_R < 0$ in the region ${\mathcal S}_1(1) \cup {\mathcal S}_2$.
Namely, the remaining task is to prove $\dot{U}_R < 0$
in the region of ${\mathcal S}_1(\varepsilon_R')\setminus{\mathcal S}_2$.

Equation (\ref{eq:dotUR2}) is also rewritten as
\begin{eqnarray}
U_R &\leq& k_e \rho_R + \sum_{i \in \V_f} k_e \Bigl( - \phi(e^{-\hat{\xi}\theta^*} e^{\hat{\bar{\xi}}\bar{\theta}_{w,i}}) \nonumber \\
&& \hspace{1cm}- \beta (1 - \epsilon) \phi(e^{-\hat{\bar{\xi}}\bar{\theta}_{w,i}} e^{\hat{\xi}\theta_{wo_i}})\Bigr) - a_R,
\label{eq:urdot3} \\
a_R &:=& \beta \Bigl( \sum_{i \in \V_f}k_e \epsilon \phi(e^{-\hat{\bar{\xi}}\bar{\theta}_{w,i}} e^{\hat{\xi}\theta_{wo_i}}) \nonumber \\
&& \hspace{1.5cm}+ \sum_{i \in \V}\sum_{j \in \N_i} k_s \phi(e^{-\hat{\bar{\xi}}\bar{\theta}_{w,i}} e^{\hat{\bar{\xi}}\bar{\theta}_{w,j}}) \Bigr),
\label{eq:ar}
\end{eqnarray}
where $a_R$ is strictly positive under Assumption \ref{assum:static}.
Now, for any $a \in (0,1)$ and $j^* \in \V$, we have
\begin{eqnarray}
\phi(e^{-\hat{\bar{\xi}}\bar{\theta}_{w,i}} e^{\hat{\xi}\theta_{wo_i}}) &\geq& a \phi(e^{-\hat{\bar{\xi}}\bar{\theta}_{w,j^*}} e^{\hat{\xi}\theta_{wo_i}}) \nonumber \\
&& - \frac{a}{1 - a} \phi(e^{-\hat{\bar{\xi}}\bar{\theta}_{w,j^*}}e^{\hat{\bar{\xi}}\bar{\theta}_{w,i}}) .
\label{eq:sqare_R}
\end{eqnarray}
Let $j^*$ be a node satisfying $j^* = \arg \min_{i_0}D(i_0)$ and $G^*_T = ({\cal V}{\cal E}^*_T) \in {\cal T}(j^*)$ be the graph satisfying $G^*_T = \arg \min_{G_T \in {\cal T}(j^*)}\tilde{D}(G_T)$.
Then, we obtain
\begin{eqnarray}
&&\phi(e^{-\hat{\bar{\xi}}\bar{\theta}_{w,j^*}}e^{\hat{\bar{\xi}}\bar{\theta}_{w,i}}) \leq \nonumber \\
&&d_{G^*_T}(i) \sum_{l \in \{ 0, \cdots, d_{G^*_T}(i)-1 \}} \phi(e^{-\hat{\bar{\xi}}\bar{\theta}_{w,v_l(i)}}e^{\hat{\bar{\xi}}\bar{\theta}_{w,v_{l+1}(i)}}), \nonumber 
\label{eq:pathR}
\end{eqnarray}
where $(v_0(i), \cdots, v_{d_{G^*_T}(i)-1}(i))$ is the path from root $j^*$ to node $i$ along the tree $G_T^*$.
Namely, 
\begin{eqnarray}
&&\sum_{i \in \V_f} \phi(e^{-\hat{\bar{\xi}}\bar{\theta}_{w,j^*}}e^{\hat{\bar{\xi}}\bar{\theta}_{w,i}}) \leq  \nonumber \\
&&\sum_{i \in \V_f} d_{G^*_T}(i) \sum_{l \in \{ 0, \cdots, d_{G^*_T}(i)-1 \}} \phi(e^{-\hat{\bar{\xi}}\bar{\theta}_{w,v_l(i)}}e^{\hat{\bar{\xi}}\bar{\theta}_{w,v_{l+1}(i)}}) \nonumber 
\label{eq:pathR1}
\end{eqnarray}
holds.
For any edge $E = (v^1, v^2)$ of $G_T^*$, the coefficient of 
 $\phi(e^{-\hat{\bar{\xi}}\bar{\theta}_{w,v^1}}e^{\hat{\bar{\xi}}\bar{\theta}_{w,v^2}})
 $ in the right hand side of (\ref{eq:pathR1}) is
given by $\sum_{i \in \V_f} \Delta_{G_T^*}(E;i) d_{G^*_T}(i)$, which is upper-bounded by $\tilde{D}(G^*_T) = W$.
We thus have
\begin{eqnarray}
\sum_{i \in \V_f} \phi(e^{-\hat{\bar{\xi}}\bar{\theta}_{w,j^*}}e^{\hat{\bar{\xi}}\bar{\theta}_{w,i}}) &\leq& W \sum_{E = (v^1, v^2) \in \E_T} \phi(e^{-\hat{\bar{\xi}}\bar{\theta}_{w,v^1}}e^{\hat{\bar{\xi}}\bar{\theta}_{w,v^2}}) \nonumber \\
&\leq& W \sum_{i \in \V}\sum_{j \in \N_i} \phi(e^{-\hat{\bar{\xi}}\bar{\theta}_{w,i}}e^{\hat{\bar{\xi}}\bar{\theta}_{w,j}}).
\label{eq:pathRW}
\end{eqnarray}
The latter inequality of (\ref{eq:pathRW}) holds because $G^*_T$ is a subgraph of $G$. 

Suppose that $(e^{\hat{\bar{\xi}}\bar{\theta}_{w,i}})_{i \in {\cal V}} 
 \in {\mathcal S}_1(\varepsilon_R')\setminus{\mathcal S}_2$.
Then, the inclusion $(e^{\hat{\bar{\xi}}\bar{\theta}_{w,i}})_{i \in \V} \notin {\cal S}_2$ holds and hence
\begin{eqnarray}
\sum_{i \in \V}\sum_{j \in \N_i} \phi(e^{-\hat{\bar{\xi}}\bar{\theta}_{w,i}}e^{\hat{\bar{\xi}}\bar{\theta}_{w,j}}) \leq \frac{k \rho_R}{\beta}.
\label{eq:nots2R}
\end{eqnarray}
From the definition of average $e^{\hat{\xi}\theta^*}$, we also have
\begin{eqnarray}
\sum_{i \in {\cal V}_f} \phi(e^{-\hat{\bar{\xi}}\bar{\theta}_{w,j^*}} e^{\hat{\xi}{\theta}_{wo_i}}) \geq \rho_R.
\label{eq:sqare_R1}
\end{eqnarray}
Substituting (\ref{eq:sqare_R}), (\ref{eq:pathRW}), (\ref{eq:nots2R}) and (\ref{eq:sqare_R1}) into (\ref{eq:urdot3}) yields
\begin{eqnarray}
\hspace{-1cm}\dot{U}_R &\leq& - k_e\sum_{i \in \V_f} \phi(e^{-\hat{\xi}\theta^*} e^{\hat{\bar{\xi}}\bar{\theta}_{w,i}}) \nonumber \\
\hspace{-1cm}&& + k_e \Big(1 - (1 - \epsilon)(a \beta - \frac{kW a}{1 - a})\Big)\rho_R - a_R.
\label{eq:urdot4}
\end{eqnarray}
Since (\ref{eq:urdot4}) holds for all $a \in (0,1)$, if we set $a = 1 - 
 \sqrt{kW/\beta}$ for a sufficiently small $k$ satisfying $1 - 
 \sqrt{kW/\beta} \in (0,1)$, we obtain
\begin{eqnarray}
\dot{U}_R &\leq& - k_e\sum_{i \in {\cal V}_f}
\phi(e^{-\hat{\xi}\theta^*} e^{\hat{\bar{\xi}}\bar{\theta}_{w,i}}) + k_e \varepsilon_R'  \rho_R - a_R. \nonumber \label{eq:urdot5} 
\end{eqnarray}
Moreover, because of $(e^{\hat{\bar{\xi}}\bar{\theta}_{w,i}})_{i \in {\cal V}} 
 \in {\mathcal S}_1(\varepsilon_R')$, 
\[
 \dot{U}_R \leq -a_R < 0
\]
holds true.
This completes the proof.
\qed
\end{pf}

We are now ready to prove Theorem \ref{thm:eps-level}.
We immediately see from Lemma \ref{lem:l2}
that the trajectories of orientation estimates $(e^{\hat{\bar{\xi}}\bar{\theta}_{w,i}})_{i \in {\cal V}}$ along with 
(\ref{eq:es_ori_f}) settle into the set 
$\Omega_R(\varepsilon)$ with $\varepsilon$ satisfying 
\begin{equation}
{\cal S}_3(\varepsilon'_R) := {\cal S}\setminus ({\cal S}_1(\varepsilon'_R) \cup {\cal S}_2)
\subseteq \Omega_R(\varepsilon).
\label{eq_hata3}
\end{equation}

Let us next derive an upper-bound of the minimal $\varepsilon$
satisfying (\ref{eq_hata3}).
For this purpose, we consider 
\begin{eqnarray}
& \max_{(e^{\hat{\bar{\xi}}\bar{\theta}_{w,i}})_{i\in \V}}& \frac{1}{n}\sum_{i \in {\cal V}} \phi(e^{-\hat{\xi}\theta^*}e^{\hat{\bar{\xi}}\bar{\theta}_{w,i}})
\label{eq:max_R} \\
& \mbox{subject to } & \sum_{i \in {\cal V}_f} \phi(e^{-\hat{\xi}\theta^*}e^{\hat{\bar{\xi}}\bar{\theta}_{w,i}}) \leq \varepsilon'_R\rho_R, \label{eq:max_R_c1} \\
&& \sum_{i \in {\cal V}}\sum_{j \in {\cal N}_i} \phi(e^{-\hat{\bar{\xi}}\bar{\theta}_{w,i}}e^{\hat{\bar{\xi}}\bar{\theta}_{w,j}}) \leq \frac{k \rho_R}{\beta}.
\label{eq:max_R_c2}
\end{eqnarray}
To compute an upper bound of the optimal value,
we relax the constraints (\ref{eq:max_R_c2}) as
\begin{eqnarray} 
\phi(e^{-\hat{\bar{\xi}}\bar{\theta}_{w,i}}e^{\hat{\bar{\xi}}\bar{\theta}_{w,j}}) \leq
\frac{k \varepsilon'_R\rho_R}{2\beta} \  \forall j \in {\cal N}_i. \label{eq:max_R1_c21} 
\end{eqnarray}
Any two nodes are connected by a path over graph $G$
whose length is smaller than the diameter of the graph $G$
denoted by ${\rm diam}(G)$.
Thus, 
(\ref{eq:max_R1_c21}) implies that
\begin{eqnarray}
\!\!\!\!\!\!\phi(e^{-\hat{\bar{\xi}}\bar{\theta}_{w,i}}e^{\hat{\bar{\xi}}\bar{\theta}_{w,j}}) \leq
\alpha_R:=
\frac{k \rho_R {\rm diam}(G)}{2\beta} \  \forall i,j \in \V. 
\label{eq:max_R2_c2}
\end{eqnarray}
Note that $\lim_{k\rightarrow 0}\alpha_R = 0$.
If we define $R_i = I_3 - e^{-\hat{\xi}\theta^*}e^{\hat{\bar{\xi}}\bar{\theta}_{w,i}}$,
the problem (\ref{eq:max_R}), (\ref{eq:max_R_c1}), (\ref{eq:max_R2_c2})
is rewritten as
\begin{eqnarray}
& \max_{(R_i)_{i\in \V}} & \frac{1}{2n}\sum_{i \in \V} 
\|R_i\|_F^2  \label{eq:max_R3} \\
& \mbox{subject to }& \frac{1}{2}\sum_{i \in \V_f} \|R_i\|_F^2 \leq \varepsilon'_R\rho_R,  \label{eq:max_R3_c1}\\
&& \frac{1}{2}\|R_i - R_j\|_F^2 \leq \alpha_R \  \forall i,j \in {\cal V}.
\label{eq:max_R3_c2}
\end{eqnarray}
For any $i\in \V$ and $j \in \V_f$, (\ref{eq:max_R3_c2})
implies that 
\[
\|R_i\|_F \leq \|R_i - R_j\|_F + \|R_j\|_F \leq \sqrt{2\alpha_R} + \|R_j\|_F.
\]
Now, it is clear that the optimal solution to
(\ref{eq:max_R3}) has to satisfy $\|R_i\|_F \neq 0\ {\forall i}\in \V$
and there exists a sufficiently small $k$ such that 
$\|R_i\|_F - \sqrt{2\alpha_R} > 0$.
We thus obtain
\[
\frac{1}{2}|\V_f|(\|R_i\|_F-\sqrt{2\alpha_R})^2 \leq 
\frac{1}{2}\sum_{j \in \V_f}\|R_j\|^2_F \leq \varepsilon'_R\rho_R
\]
from (\ref{eq:max_R3_c1}) and hence
\[
 \|R_i\|_F \leq \sqrt{2}(\sqrt{\alpha_R} + \sqrt{\varepsilon'_R\rho_R/|\V_f|}).
\]
Namely, the optimal value of (\ref{eq:max_R3}) 
is upper bounded by
$(\sqrt{\alpha_R}+\sqrt{\varepsilon'_R\rho_R\|\V_f|\|})^2$ and hence $\varepsilon$
is also bounded by $(\sqrt{\alpha_R|\V_f|/\rho_R}+\sqrt{\varepsilon'_R})^2$.
Since $\lim_{k\rightarrow 0}\alpha_R = 0$ and $\lim_{k\rightarrow 0}\varepsilon_R' = \varepsilon_R$,
\[
\lim_{k\rightarrow 
 0}(\sqrt{\alpha_R|\V_f|/\rho_R}+\sqrt{\varepsilon'_R})^2 = 
 \varepsilon_R 
\]
holds.
This completes the proof.

\end{document}